 \documentclass[rnote]{aa} 
\usepackage{graphicx}
\usepackage{txfonts}
\usepackage{natbib}
\begin{document}
   \title{Hard X-ray flux from low-mass stars in the
\object{Cygnus~OB2} Association}

\author{M.~Caramazza\inst{1,2}, J.J.~Drake\inst{3}, G.~Micela\inst{2,4},
E.~Flaccomio\inst{2,5}}

\institute{Dipartimento di Scienze Fisiche ed Astronomiche,
Universit\`a di Palermo, Via Archirafi 36, 90123 Palermo, Italy
\and
INAF Osservatorio Astronomico di Palermo, Piazza del Parlamento 1,
90134 Palermo, Italy \\
\email{mcarama@astropa.unipa.it}
\and
Smithsonian Astrophysical Observatory, MS-3, 60 Garden Street,
Cambridge MA~02138\\
\email{jdrake@cfa.harvard.edu}
\and
\email{giusi@astropa.unipa.it}
\and
\email{flaccomio@astropa.unipa.it}
}

   \date{Received; accepted }


  \abstract
{The Cygnus~OB2 association, the central engine of the Cygnus X star-forming
region, is the subject of an extensive INTEGRAL Key Project
that will accumulate 6Ms of observations. Analysis of 2Ms of observations
by De~Becker and co-workers provides the most sensitive limit yet obtained
on hard X-ray emission from the cluster.
}
   {We investigate the X-ray emission in the 20-40~keV band
expected from the flaring low-mass stellar population in
Cygnus~OB2.  We discuss whether such emission needs to be
considered in the interpretation of existing and future X-ray observations of
the region, and whether such observations might provide insight into the
high-energy processes on low-mass pre-main sequence stars.}
{The total hard X-ray flux from low-mass stars
is estimated by assuming the observed soft X-ray emission
stems from a superposition
of flares.  We further
assume the ratio of hard X-ray to soft X-ray emission is
described by a scaling found for solar flares by Isola and co-workers.}
{We estimate the low-mass stellar hard X-ray flux in
the 20-40~keV band to lie in the range
{$ \sim2 \times 10^{31}-6 \times 10^{32} \ \ erg\ s^{-1}$}  and
discuss some potential biases that might affect this result.}
   {Hard X-ray emission could lie at a level not much below the current
observed flux upper limits for Cygnus~OB2.  If this emission could be
detected, it would provide insight into the hard X-ray production
of large flares on pre-main sequence stars. We highlight the penetrating
power of hard X-rays from low-mass stellar populations as a possible pointer
to our Galaxy's hidden star-forming clusters and super-clusters using
more sensitive observations from future missions.}

\keywords{Stars:coronae - flare -
                Open Cluster and Association: Cygnus~OB2      }

\titlerunning{Hard X-rays from low-mass stars in Cyg OB2}
\authorrunning{M.~Caramazza et al.}
\maketitle


\section{Introduction}

Motivated by the detection of nonthermal radio emission from
early-type stars in the nearby massive Cygnus~OB2 association
\citep[$d=1.7$~kpc][]{Massey.Thompson:91}, \citet{De_Becker.etal:07}
have recently used an extensive set of observations obtained by the
International Gamma-Ray Laboratory (INTEGRAL) IBIS instrument to
search for hard X-ray emission from some of the more prominent radio
sources.  The nonthermal radio emission from these stars is thought
to arise from a relativistic population of electrons accelerated by
diffusive shock acceleration in their interacting stellar winds
\citep[e.g.][]{Pittard.Dougherty:06}.  \citet{De_Becker.etal:07} note
that hard X-rays might plausibly be produced by inverse Compton
scattering of photospheric UV photons.

One other potential source of hard X-rays in Cygnus~OB2 are the myriad
flares thought to arise from magnetic reconnection events that
characterize and sustain the coronae of the low-mass stellar
population \citep[see, e.g.,][]{Albacete_Colombo.etal:07, Caramazza.etal:07}. 
The flare
paradigm developed from the solar perspective posits that hard X-rays
can arise from the impact on dense photospheric gas of electrons
accelerated to mildly relativistic energies \citep[the ``thick-target
model'', e.g.][]{Brown:71,Hudson:72,Lin.Hudson:76}.  Tentative direct
evidence for such nonthermal hard X-rays up to 200~keV has recently
been found from {\it Swift} observations of a large flare on the
RS~CVn-like interacting binary II~Peg \citet{Osten.etal:07}.  Large
stellar flares observed on the young single star AB~Dor and the active
binaries Algol and UX~Ari by the {\it Beppo}SAX mission
\citep{Boella.etal:97} in the 0.1-100~keV range revealed evidence of
similarly energetic emission, though it was not possible to ascertain
whether this was of thermal or nonthermal origin
\citep[][]{Maggio.etal:00,Favata.Schmitt:99, Franciosini.etal:01}.

While any single flare event on a low-mass star is a trifling
insignificance in comparison with the effusive vigor of its high-mass
brethren, the comparative multitude of the former population in
Cygnus~OB2 might render a palpable combined effect.  An estimate of
this effect is motivated from different standpoints.

Firstly, the search
for hard X-rays from early-type stars by \citet{De_Becker.etal:07}
resulted in only upper limits; an estimate of the flux limit at which
low-mass stars might be detected could be important for understanding
the origin of any hard X-rays detected in deeper observations by
INTEGRAL or future missions.

Secondly, the large concentration of low-mass stars in the Cygnus~OB2
region might also provide insight into the high-energy nature of
stellar flares that is difficult to attain from observations of nearby
single stars that rarely exhibit large events.  \citet{Isola.etal:07}
have recently shown that a well-defined power-law relationship between
soft and hard X-rays seen in solar flares matches the 20-40~keV
flux observed during the AB~Dor, Algol and UX~Ari flares, but
tends to overpredict the, presumed nonthermal,
60-80~keV flux. \citet{Alexander.Daou:07} find evidence for a saturation in
nonthermal X-ray flux above 20~keV for solar flares with increasing
soft X-ray luminosity, and it is conceivable that this is related to the
deficit in observed nonthermal stellar X-rays.
Detection of, or limits on, the hard X-ray flux
from the low-mass Cygnus~OB2 population could provide constraints for
extrapolating solar flare models to much more active pre-main sequence
stars.

Thirdly, the hard X-ray environment of young protostellar
systems is potentially relevant for the ionization of
circumstellar material and protoplanetary disks
\citep[e.g.][]{Glassgold.etal:97,Aikawa.Herbst:99,Fromang.etal:02,
Matsumura.Pudritz:03}.  We currently have no information on the intensity
of this radiation environment.

We outline in Sect.~\ref{s:estimate} below a calculation of the hard X-ray flux
from the
Cygnus~OB2 low-mass population under the assumption that the observed
soft X-rays from these stars is dominated by a superposition of
flaring events.  We then discuss in Sect.~\ref{s:discuss} the
uncertainties in this estimate, the degree to which it likely
overestimates the true level of hard X-ray emission, and the
implications of the results.

\section{Estimating the low-mass stellar hard X-ray flux}
\label{s:estimate}

There are three main components to our hard X-ray flux estimate: the
Cygnus~OB2 initial mass function (IMF) and total stellar mass; the
relationship between stellar mass and observed soft X-ray luminosity
for the low-mass population; and the relationship between the soft and
hard X-ray luminosities.  In order to gain a better perspective on the
uncertainties involved in our estimate, we use three different approaches
that employ these ingredients in slightly different ways.

We adopt the IMF derived by \citet{Knodlseder:00} based on 2MASS
photometry.  That study noted the power-law index found,
$\Gamma=-1.6\pm0.1$, is compatible with that of \citet{Kroupa.etal:93}
for the solar neighborhood, but is significantly steeper than found
from spectroscopy of the central region of Cygnus~OB2 by
\citet{Massey.Thompson:91}.  We therefore also investigate below the
effect of a different power-law slope, and in particular that found
for the Orion Nebula Cluster (ONC; $\Gamma=-1.2$) by
\citet{Muench.etal:02}.  We
consider hard X-rays from stars in the mass range 0.3--$2M_\odot$; the
upper limit corresponds to the approximate boundary between low-mass
pre-main sequence stars with outer convection zones and intermediate
mass stars that should have radiative envelopes; the lower mass limit
corresponds to the limit of the \citet{Knodlseder:00}
study---essentially our limit of current knowledge.  For stars with
mass $\la 0.3M_\odot$, X-ray luminosities of T Tauri stars are observed to
decline much more rapidly with decreasing mass than for higher mass stars
\citep[e.g.][]{Preibisch.etal:05,Albacete_Colombo.etal:07b}, rendering
their contribution to the total X-ray luminosity less significant,
regardless of whether the IMF flattens toward lower masses or not
\citep[e.g.][]{Scalo:86,Kroupa.etal:93,Kroupa:01,Chabrier:03}.
Nevertheless, we also investigate the effects of this lower mass
cut-off in our calculations below.

\citet{Albacete_Colombo.etal:07} found the X-ray luminosities of
low-mass stars in Cygnus~OB2 to be essentially identical to those in
the ONC \citep{Preibisch.etal:05}.  We therefore adopt the relation between
stellar mass,
$M$, and X-ray luminosity, $L_X$, from the ONC study. In order to examine the
effects on our
calculation of the significant scatter in observed $L_X$ vs $M$, we also
calculate
the expected hard X-ray flux using the observed ONC stellar sample
renormalized to the Cygnus~OB2 low-mass content (see Sect.~\ref{s:scale}).

The relationship between the soft and hard X-ray luminosities was
derived using the extrapolation of the relation found for solar flares
by \citet{Isola.etal:07}, $F_{20-40}=9.9\times 10^6 F^{1.37}_G$, where
$F_{20-40}$ is the 20-40~keV  flux density in ph~cm$^{-2}$~keV$^{-1}$~s$^{-1}$
and $F_G$ is
the {\it Geostationary Operational Environment Satellite} (GOES)
1.6--12.4~keV flux in unit of W~m$^{-2}$.  In order to convert this relation to
the stellar
case, we assumed that the observed soft X-ray luminosities of the
Cygnus~OB2 low-mass pre-main sequence population could be explained by
a superposition of flares.  This assumption is based on evidence that
the solar corona is largely characterized by a power-law flare
distribution in total energy, $E$, and frequency, $N$, $dN/dE\propto
E^{-\alpha}$ \citep[e.g.][]{Lin.etal:84,Krucker.Benz:98,Hudson:91},
and on studies of EUV and X-ray flares and photon arrival times for
active stars that suggest a similar flare distribution with $\alpha
\sim2$ \citep{Audard.etal:00,Kashyap.etal:02,Gudel.etal:03}.  The
high-energy tails of flares detected in {\it Chandra} studies of the
ONC and Cygnus~OB2 are also well-approximated by a power-law frequency
distribution with $\alpha=2.2 \pm 0.2$ \citep[ONC,][]{Caramazza.etal:07},
$\alpha=2.1 \pm 0.1$ \citep[Cygnus~OB2,][]{Albacete_Colombo.etal:07}.
In the following we outline our estimates, and then discuss the uncertainties of
this approach. An alternative approach to estimate the nonthermal emission
is proposed in \citep{Guedel:09}.


%
\subsection{Hard-soft X-ray luminosity relation for young low-mass stars}
\label{star_relation}
The relation from \citet{Isola.etal:07} relates the flux in the $1.6-12.4$ keV
band to the flux density in the $20-40$ keV band. In order to obtain a relation
between the peak luminosity of flares in the two bands, we first converted the
GOES flux from  W~m$^{-2}$ into erg s$^{-1}$ luminosity, and the  flux density
in the hard band, given in unit of ph~cm$^{-2}$~keV$^{-1}$~s$^{-1}$, into erg
s$^{-1}$ luminosity, assuming a photon energy of 20 keV and multiplying for the
band range. We also converted the $1.6-12.4$ keV luminosity into luminosity in
the $0.5-8.0$ keV band: the conversion factor ($K_{conv}=0.34$) was evaluated
considering a Raymond Smith model with $kT=1.35$ keV and $log(N_H)=22.25$
cm$^{-2}$ \citep{Albacete_Colombo.etal:07b}. We then obtain the following
relation:
\begin{equation} \label{e:isola}
\mathcal{L}_{fl(20-40)}=1.8 \cdot 10^{28} \cdot \left( 1.2 \cdot
10^{-31}\right)^\beta \cdot \left(\mathcal{L}_{fl(0.5-8.0)}\right)^{\beta} \ \
erg \ s^{-1}
\end{equation}

where {$\beta=1.37 \pm 0.07$ is the index found by  \citet{Isola.etal:07}. Note
that the factor in the previous relation varies strongly due to the uncertainty
of $\beta$.}

 This relation can be converted in a relation between the total energy output of
 flares in the two spectral bands, considering that $E=\mathcal{L}_{fl} \cdot
 \tau$. Given the power-law distribution of flare energies in the $0.5-8.0$ keV
band, we conclude that the $20-40$ keV band flare energies are also distributed
as a power-law.  The index of this
power-law, $\gamma$, is a function of $\alpha$ and $\beta$:
$\gamma=(\alpha+\beta-1)/\beta$, where $2.0 \leq \alpha \leq 2.4$
\citep{Caramazza.etal:07} and $1.30 \leq \beta \leq 1.44$ \citep{Isola.etal:07}.
 If we impose the condition that
$\gamma$ also has a value greater than $2$, i.e. that the $20-40$ keV
band emission is due entirely to flares, then the acceptable values
for $\alpha$ and $\beta$ fall in the ranges $1.31<\beta<1.39$ and
$2.32<\alpha<2.4$. Integrating the distribution of flares between the
minimum energy and the
maximum energy, that we can set as infinite, we obtain the following
relation between stellar luminosity in the $20-40$ keV and $0.5-8.0$ keV
bands:
\begin{equation} \label{eq:hard_source}
L_{20-40}=A \cdot L_{0.5-8.0}\ \
erg \ s^{-1}
\end{equation}
where
\begin{equation}
A=\frac{\tau_{20-40}}{\tau_{0.5-8.0}^\beta} \frac{(\alpha-2) \cdot E_{0.5-8.0
min}^{\beta -1}}{\alpha-\beta-1} 1.8 \cdot 10^{28} \cdot \left( 3.5 \cdot
10^{-31}\right)^\beta
\end{equation}
Here, we set $\tau_{0.5-8.0}$ to 10 hours \citep{Caramazza.etal:07}, while
$\tau_{20-40}$ is scaled using the ratio (0.16) derived from the median duration
of flares in the hard and soft bands calculated
by \citet{Veronig.etal:02b} from a sample of solar flares. $E_{0.5-8.0 min}$ was
set to $\sim2 \cdot 10^{32}$ erg that is the median values obtained from a
statistical analysis of the ONC lightcurves interpreted assuming the emission
enterely due to flares with a power law distribution \citep{Caramazza.etal:07}.
The energy of flares that contribute to the star emission in
\citet{Caramazza.etal:07} is given in counts. We converted the counts to energy,
assuming a Raymond-Smith emission with the median values of N$_H$ and kT of the
ONC sample (N$_H=1.9\cdot 10^{21} cm^{-2}$ $kT=1.25$ keV). Note that, since
$\beta$ is not too different from 1, the hard x-ray luminosity is relatively
insensitive to change in $E_{0.5-8.0 min}$  or $\tau_{0.5-8.0}$ values: a factor
of 10 change in either of these quantities results in only a factor of 2
difference in A. Assuming uniform distributions of $\alpha$ and $\beta$ in the
 total intervals given above, we estimate that 90\% of the resulting values
for the proportionality factor A lie in the range A=[$1.8 \cdot 10^{-4}$, $3.0
\cdot 10^{-3}$]. Allowing also for significant uncertainties in the other
parameters ($\tau_{20-40}/\tau_{0.5-8.0}$=[0.08, 0.24], $\tau_{0.5-8.0}$=[3, 18]
hours, $E_{0.5-8.0 min}$=[$2 \cdot10^{31}, 2 \cdot10^{33}$] erg), the 90\%
interval for A becomes [$1.1 \cdot 10^{-4}$, $4.0 \cdot 10^{-3}$]. We will adopt
in the following the latter uncertainty range.

\subsection{Analytical Estimate}
\label{s:analytic}

Adopting the relation between stellar mass and the X-ray luminosity
found for the {\it Chandra Orion Ultra-deep Project} (COUP) stars in
the range $0.3-2 M_{\odot}$ \citep{Preibisch.etal:05},
\begin{equation}
\log \left(L_{0.5-8.0}\right)=30.37+1.44 \cdot \log\left(M\right)
\label{e:preib}
\end{equation}

and combining this with
Eqn.~\ref{eq:hard_source}, we obtain a relation between stellar mass
and hard X-ray flux. This expression can then be integrated over the
stellar mass distribution for Cygnus~OB2.  The IMF found by
\citep{Knodlseder:00} follows the standard form
\begin{equation} \label{eq:IMF_cygnus}
\frac{dN}{dM}=k M^{\Gamma-1}= k \cdot M^{-2.6}
\ \ \ \ {\rm with} \ \ \Gamma=-1.6,
\end{equation}
and assuming for the present a total mass of $M_{tot}=5 \cdot 10^4 M_{\odot}$
\citep{Knodlseder.etal:00},
and maximum and minimum stellar masses in the cluster of
$M_{max}=80 M_{\odot}$ and $M_{min}=0.3 M_{\odot}$, respectively,
the normalization constant is $k=1.5 \cdot  10^4$.

In order to calculate the total hard X-ray luminosity in the cluster,
we transform the IMF into a function of 20-40~keV stellar X-ray
luminosity, and integrate between the luminosities corresponding to
$0.3$ and $2.0 M_{\odot}$, obtaining values in the following range:
 \begin{equation} \label{eq:calc_result}
L_{20-40}^{Cyg}=[1.8 \cdot 10^{31},6.6 \cdot 10^{32}]\ erg \ s^{-1}
\end{equation}
\subsection{Scaling the Orion Hard X-ray luminosity}
\label{s:scale}

We can also make a rough estimate of hard X-ray flux from Cygnus~OB2
by scaling directly the observed X-ray luminosity distribution of the
low-mass stars in Orion.  This approach has the advantage that the
observed scatter in the COUP sample is intrinsically included.  The
IMF of the Orion sample has a shallower slope than that for Cygnus~OB2
of \citet{Knodlseder:00}, with $\Gamma=-1.2$ \citep{Muench.etal:02}, and we
investigate the influence of this difference in Sect.~\ref{s:simul}
below.

The total 20-40~keV luminosity for the COUP sample is
simply the sum over all stars of mass $M_i$ in the range 0.3-$2M_\odot$,
$$L_{20-40}^{COUP}=A \cdot \sum_{M_i=0.3}^{2}L_{0.5-8.0}(i)=
[8.6\cdot 10^{28},3.1\cdot 10^{30}] \ \ erg \ s^{-1},$$
where the relation between the 20-40~keV and 0.5-8~keV luminosities
is given by Eqn.~(\ref{eq:hard_source}).  Assuming a similar IMF for both
Orion and Cygnus~OB2, the hard X-ray luminosity of the latter is then
simply given by the product of $L_{20-40}^{COUP}$ and the ratio
of cluster masses within the 0.3-$2M_\odot$ mass interval.
The Cygnus~OB2 IMF from Eqn.~\ref{eq:IMF_cygnus} yields a total
mass in this range of $M_{0.3-2}^{Cyg}=3.5 \cdot 10^4 M_{\odot}$,
while the analogous total mass of the COUP sample is $M_{0.3-2}^{COUP}=175
M_{\odot}$. {Even in this case the total Cygnus~OB2 hard X-ray luminosity is
therefore $$L_{20-40}^{Cyg}=[1.7\cdot 10^{31},6.3\cdot 10^{32}] \ erg \ s^{-1}$$}.
%

\subsection{Simulating Cygnus~OB2 using the Orion sample}
\label{s:simul}

In order to account for the different IMF slopes in Cygnus~OB2 and
Orion ($\Gamma=-1.6$ {\it cf} $-1.2$), we also performed a simple Monte
Carlo simulation of 10000 stars.  The mass range $0.3-2 \ M_{\odot}$
was divided into 10 bins, and within each mass bin $0.5-8.0$~keV
luminosities were randomly selected from those observed in the COUP
sample.  The number of ``stars'' drawn in each bin was weighted
according to the Cygnus~OB2 IMF from Eqn.~(\ref{eq:IMF_cygnus}), and the
luminosity for each star was scaled to $L_{20-40}$ as described
earlier.  The luminosity of the simulated sample was then scaled so as
to have a total mass equal to that of Cygnus~OB2 for the
0.3-$2M_\odot$ mass range.  In this way, we estimate
$$L_{20-40}^{Cyg}=[1.9\cdot 10^{31},6.6\cdot 10^{32}] \ erg \ s^{-1}.$$

As a verification of the calculation, we recovered a luminosity of
$L_{20-40}^{COUP}=[8.9\cdot 10^{28},3.2\cdot 10^{30}] \ erg \
s^{-1}$---essentially the
same as that found from direct calculation in Sect.~\ref{s:scale}---for
an IMF with slope $\Gamma=-1.2$ scaled to the appropriate COUP total
mass.

\section{Discussion}
\label{s:discuss}
The range of 20-40~keV fluxes we have evaluated above can be compared with the
the $3\sigma$ 20-60~keV flux upper limit of $6.1\times 10^{-12}\ erg \ cm^2\
s^{-1}$
obtained from 2.12~Ms of {\it INTEGRAL} IBIS-ISGRI observations by
\citet{De_Becker.etal:07} for the unidentified $\gamma$-ray source
3EG~J20033+4188 that lies in the close vicinity of
Cygnus~OB2 \footnote{The distance between the position of
  3EG~J20033+4188 and the center of Cyg OB2 obtained by
  \citep{Knodlseder.etal:00} is $7 \arcmin$}. Adopting a distance of
1.7~kpc for Cygnus~OB2 \citep[e.g.][]{Massey.Thompson:91}, this flux
corresponds to a luminosity $\sim2\times 10^{33}$~erg~s$^{-1}$.

The 20-40~keV bandpass considered here is significantly narrower that
the 20-60~keV range cited by \citet{De_Becker.etal:07}. Based on the
power-law spectrum and slope estimate by \citet{Isola.etal:07}, we
expect the corresponding estimate for the 20-60~keV range to be a
factor 1.8 higher.  Moreover, the upper limit of the fluxes for the
undetected stars was estimated by \citet{De_Becker.etal:07} within a
PSF of $12 \arcmin$.  Using the King profile describing the projected spatial
distribution of starlight from the association
found by \citep{Knodlseder.etal:00}, the ratio between the flux in the
whole association area (radius $\sim60 \arcmin$) to that in the
3EG~J20033+4188 region is $6.6$. Applying these scaling factors, the
total luminosity in the whole region of Cygnus should not be higher
than $7.3 \times 10^{33}$~erg~s$^{-1}$. While this scaling is somewhat
crude, the range of 20-40~keV fluxes we have evaluated is one order of
magnitude lower than this upper limit.  This confirms the potential importance
of hard X-ray emission from the low-mass stars in stellar clusters.

There are several assumptions that might have led us to overestimate the
Cygnus~OB2 luminosity.  First, the total
cluster mass of (4--10)$\times 10^4 M_\odot$ derived by
\citet{Knodlseder:00} has been challenged by \citet{Hanson:03}, who
estimated the cluster to lie slightly closer at 1.4~kpc and suggested
a total cluster mass closer to $10^4 M_\odot$.  Nevertheless, based on
a study of A-stars in the Cygnus~OB2 field, \citet{Drew.etal:08} find
that ``a total mass of $30\,000-40\,000 M_\odot$ would not be
surprising''. Our adopted $5\times 10^4 M_\odot$ is then perhaps up to
1.2-1.7 times too high.

We have also assumed that all of the observed stellar X-ray flux is
due to continuous flaring.  Some fraction of the observed luminosity
of these stars might also be attributed to a quiescent component.

One additional factor of uncertainty is related to the duration of flares.
Indeed, the hard X-rays seen in solar
flares that are usually associated with the flare impulsive phase
generally decay on a more rapid timescale than soft X-rays
\citep[e.g.][and references therein]{Dennis.Zarro:93,Benz:08}. We considered a
ratio between the duration of flares in hard and soft band derived from solar
measures of flare duration \citep{Veronig.etal:02b}.The case of large stellar
flares is not so clear, however.  The decay of the 14-40~keV flux in
the large II~Peg flare observed by {\it Swift} was not obviously
shorter than that for the soft X-rays.  It should also be noted that
in large stellar flares there will be a larger thermal contribution in
the 20-40~keV range than in much less energetic solar flares, in
alignment with the general correlation of increasing plasma
temperature with flare total energy \citep[e.g.][]{Feldman.etal:95}. Moreover,
also the minimum energy of the distribution of flares is uncertain, we set it to
the median value obtained from the analysis of the COUP low mass samples. In our
estimation we considered that it can be uncertain of two orders of magnitude,
that
implies a variation of a factor $\sim 5$ in the resulting total energy: note,
however, that \citet{Guedel:09} found a value even lower (few times $10^{30}$).
%

Our assumptions could lead to an overestimation of the real value of
the total luminosity, but a value of $\sim6 \cdot 10^{32}\ erg\ s^{-1}$, an
order of magnitude below the \citet{De_Becker.etal:07} flux upper
limit, would render the low-mass population of importance in searches
for emission from specific suspected sources of hard X-rays, such as
the unidentified $\gamma$-ray source 3EG~2033+4118, the unidentified
TeV source TeV~J2032+4130, and massive colliding-wind binaries.

The possibility of detecting hard X-ray emission from nearby clustered
low-mass pre-main sequence populations provides a promising means of
investigating high-energy processes on stars that are generally too
distant to study in detail individually.  Of particular interest is
the relation between soft and hard X-ray emission of solar flares
\citep[e.g.][]{Isola.etal:07,Battaglia.etal:05}, and how the underlying
physical basis can be extrapolated to the other regimes, such as
the pre-main sequence stellar case where hard X-rays from flares can
be important agents of protoplanetary disk ionisation
\citep[e.g.][]{Igea.Glassgold:99,Glassgold.etal:04}.

Our discussion assumes that all stars will be ``observed'' in the
20-40~keV range, {\it i.e.,} that X-rays will penetrate both the
circumstellar and ambient cluster and line-of-sight extinction.
Cygnus~OB2 is located behind the Great Cygnus Rift that leads to
extinction of up to (and possibly beyond) $A_V\sim10$
\citep[e.g.][]{Massey.Thompson:91}.  Extinction of $10^m$ corresponds
to a neutral hydrogen column density $N_H\sim1.5\cdot 10^{22}\ cm^{-2}$.
The optical depth for such a column for 20-40~keV X-rays
is completely negligible; the Compton ionization and scattering
cross-section of $\sigma_T\sim6.65 \cdot 10^{-25} \ cm^2$ renders the
interstellar medium optically thin to hard X-rays for columns up to
$N_H\sim10^{24} \ cm^{-2}$.  Except perhaps in rare cases of stars
obscured by very dense circumstellar disk mid-plane gas and dust, the
assumption that we will see the whole cluster in hard X-rays should be
valid.

Finally, this consideration raises the possibility of using the
penetrating power of hard X-rays as a pointer to our Galaxy's hidden
superclusters.  \citet{Hanson:03} noted that extrapolation of the
locally-derived Galactic cluster luminosity function indicates that
our Galaxy hosts ``tens to perhaps a hundred'' massive clusters with
total mass $\sim10^4 M_\odot$. These clusters are likely hidden
behind many magnitudes of extinction and will not be easy to locate.
Hard X-rays can penetrate such extinction, and the arrival of missions
in the next decade able to provide relatively precise imaging with
much greater sensitivity in the $\sim10-100$~keV bandpass, such as
{\it NuSTAR} \citep{Harrison.etal:05} and {\it Symbol-X}
\citep{Pareschi.Ferrando:05}, could provide an assay of this hitherto
much overlooked population of our Galaxy.  In this context, Cygnus~OB2
represents a potential ``Rosetta Stone'', offering a nearby
super-cluster example that can be well-characterized using
multi-wavelength techniques that will not be applicable to more
distant and extinguished clusters.

\begin{acknowledgements}

JJD was funded by NASA contract NAS8-39073 to the {\it Chandra X-ray
Center} (CXC) during the course of this research and thanks the CXC
director, Harvey Tananbaum, and the CXC science team for advice and
support.  JJD also thanks the ISHERPA program for financial support
during his visit to the Osservatorio Astronomico di Palermo, and the
Osservatorio director, Prof.~S.~Sciortino, and staff for their help
and warm hospitality.  MC, GM and EF acknowledge financial support from
the Ministero dell'Universit\`a e della Ricerca and ASI/INAF Contract
I/023/05/0.  JJD thanks J\"urgen Kn\"odlseder for useful discussions
that partly inspired this work.\\
The authors thank the referee Manuel G\"udel for suggestions and useful
comments that improved this work.
\end{acknowledgements}

\bibliographystyle{aa}
\bibliography{cygob2hard}


\end{document}